# Superconducting routing platform for large-scale integration of quantum technologies

C. Thomas[1,*], J-P. Michel[1], E. Deschaseaux[1], J. Charbonnier[1], R. Souil[1],
E. Vermande[1], A. Campo[1], T. Farjot[1], G. Rodriguez[1], G. Romano[1], F. Gustavo[2],
B. Jadot[1], V. Thiney[1], Y. Thonnart[3], G. Billiot[1], T. Meunier[4] and M. Vinet[1]

[1] Univ. Grenoble Alpes, CEA, LETI, 38000 Grenoble, France
[2] Univ. Grenoble Alpes, CEA, IRIG, 38000 Grenoble, France
[3] Univ. Grenoble Alpes, CEA, LIST, 38000 Grenoble, France
[4] Univ. Grenoble Alpes, CNRS, Néel Institute, 38000 Grenoble, France

*E-mail: candice.thomas@cea.fr

**Abstract**

To reach large-scale quantum computing, three-dimensional integration of scalable qubit arrays and their control electronics in multi-chip assemblies is promising. Within these assemblies, the use of superconducting interconnections, as routing layers, offers interesting perspective in terms of (1) thermal management to protect the qubits from control electronics self-heating, (2) passive device performance with significant increase of quality factors and (3) density rise of low and high frequency signals thanks to minimal dispersion. We report on the fabrication, using 200 mm silicon wafer technologies, of a multi-layer routing platform designed for the hybridization of spin qubit and control electronics chips. A routing level couples the qubits and the control circuits through one layer of $Al_{0.995}Cu_{0.005}$ and superconducting layers of TiN, Nb or NbN, connected between them by W-based vias. Wafer-level parametric tests at 300 K validate the yield of these technologies while low temperature electrical measurements in cryostat are used to extract the superconducting properties of the routing layers. Preliminary low temperature radio-frequency characterizations of superconducting passive elements, embedded in these routing levels, are presented.

## 1. Introduction

The high promises of quantum computation drive important efforts to develop technologies to scale qubits. Indeed, reaching quantum supremacy regime requires tens to hundreds of errorless qubits formed by arrays of thousands of physical qubits, on which error correction codes are applied [1]. The proper operation of these qubit arrays relies on several technological aspects such as the wiring and packaging. The complexity of packaging qubits essentially lies in the large number of input and output signals to address and read, all inside a cryostat, at temperatures in between a few mK to about 1 K [2]. This motivates the development of cryo-CMOS (CMOS standing for complementary metal-oxide-semiconductor) control electronics embedded at proximity of the qubits in the cryostat [3][4][5]. It also requires the design and fabrication of multi-chip assemblies using three-dimensional (3D) architectures to integrate both the qubit and cryo-CMOS chips while favoring large scale approach with flip-chip packaging techniques [6][7]. Recently, the first fabrications of multi-chip platforms and interposers hosting superconducting [8][9] and spin qubits [10] have been reported. These platforms use interconnects made from superconducting materials, with e.g. In microballs [10] and TiN-based through silicon vias (TSVs) [9]. They also integrate Nb superconducting routing levels especially for superconducting qubits [11]. While these superconducting inter- and intra-connections are mandatory for the proper integration of superconducting qubits, they offer promising perspectives for spin qubits regarding aspects such as thermal management, passive device integration as well as scalability of low and high frequency signals.

In this letter, we report on the fabrication and characterization of an interposer designed for the integration of spin qubits and their control electronics. The architecture relies on the CEA-LETI interposer and 3D integration know-how developed especially for high performance computing and photonic applications [12][13]. Using wafer-level parametric tests at 300 K and low temperature electrical measurements of single dies, we evaluate the electronic and superconducting properties of the interposer routing levels made either in $Al_{0.995}Cu_{0.005}$, TiN, Nb or NbN. This allows validating design rules compatible with the operating conditions of spin qubits in terms of temperature, magnetic field and applied current. Based on these results, radio frequency (RF) passive components have been designed and integrated on the interposer routing levels. Preliminary low temperature high frequency measurements performed on resistors, capacitors and inductors give insights on the routing and passive properties up to a few GHz, which is the frequency range typically used to encode and read spin qubits.

## 2. Demonstrator presentation

The interposer presented in this paper is part of a complete demonstrator called QuIC$^3$ (standing for Quantum Integrated Circuits with Cryo-CMOS) which is schematically presented in Figure 1(a). QuIC$^3$ comprises one chip with arrays of spin qubits and two cryo-CMOS chips for control and read-out, which are flip-chipped and hybridized on top of a silicon-based interposer. This platform is suitable for the integration of any type of spin qubit systems that have already been demonstrated [14]. By bringing the qubit arrays and cryo-CMOS control chips together, this demonstrator aims at making a first step toward scaling up. As silicon-based spin qubits appear as the most promising solution for large scale integration, our efforts with this demonstrator focus on this technology [15][16], but hybridization of other semiconductor quantum circuits, made from e.g. GaAs, can be envisioned as testbeds for proof of concepts. Cryo-CMOS chips, connected to the qubit chip via the interposer routing lines, perform control functions such as signal multiplexing and read-out using transimpedance amplifiers [17]. Additional reflectometry circuits [18][19] (labeled as on-chip passive in Figure 1(a)), comprising resistors, inductors and capacitors, are also integrated on the interposer routing levels to provide read-out alternatives.

Compared to a standard wire-bonding approach, interposer strategy has several advantages. First, hybridization and coupling of chips made from different technological nodes and materials is possible, allowing to dissociate technological challenges and separately optimize each circuits. Then, the design flexibility of the interposer routing lines enables to reduce parasitic capacitance and inductance known to introduce noise in the qubit measurements. It permits to significantly increase the connection density while limiting cross-talk between high frequency signals using microstrip, stripline or even coplanar waveguide transmission lines on the multi-layer routing levels. These levels also facilitate the integration of embedded passive components forming e.g. read-out circuits. Moreover, the choice of materials and technologies included in the interposer provide solutions to thermally decoupled the quantum and cryo-CMOS chips, protecting the qubits from cryo-CMOS self-heating [20][21] to limit decoherence mechanisms.

As shown in Figure 1(b), on its front-side the interposer contains several routing levels, allowing the supplies of embedded on-chip passives and of the hybridized cryo-CMOS circuits as well as their electrical connection with the qubit chip. Two layers (MET1 and MET2) are connected through vias. Standard microelectronic metallic stacks such as Ti/TiN/AlCu and Ti/TiN/W can be used to form these layers and vias as well as sole type II superconducting materials such as TiN, Nb and NbN. The superconducting phase of these materials, resulting in electrical resistance vanishing and thermal resistance rise due to the formation of Cooper pairs, should increase the quality factor of the on-chip integrated passives and help thermally isolate the qubits. The choice of the superconductors relies on two aspects: the integration maturity of these materials and their promise in terms of superconducting properties. Widely used as a barrier layer in common CMOS device fabrication, TiN is nowadays a part of superconducting qubit integration as, e.g., under bump metallisation (UBM) [22], TSVs [9] and for low-loss passive elements [23]. Integration of Nb in multi-layer back end of line has been developed over the years for large-scale single flux quantum circuits [24][25] and is now the main part of routing in multi-chip modules and interposers dedicated to superconducting qubits [11]. Multi-chip modules made from NbN routing have also been previously studied for high-speed digital applications [26]. NbN is now incorporated in bumping technologies for 3D quantum architectures [27] or in qubit read-out elements such as resonators [28] thanks to its large kinetic inductance. These integration developments and studies provide a solid basis for the incorporation of these superconductors in complex 3D architectures. In the following, we will also show that industrially compatible Ti/TiN/AlCu layer stack presents interesting superconducting properties.

On the backside of the interposer, a packaging support, such as a land grid array (LGA), connects the demonstrator to a printed circuit board and the cryostat thanks to Cu-based TSVs, a Cu-based redistribution layer and SnAg microbumps. Standard conductive materials are used for these interconnects to help thermalizing the interposer. The choice of SnAg-based microbumps in between the interposer and the LGA relies on early-on studies showing evidences of their mechanical and electrical reliability at sub-Kelvin temperatures [29] [30].

In the following, we will focus on the interposer part of QuIC$^3$ demonstrator, more specifically on the fabrication and characterizations of the front-side routing levels.

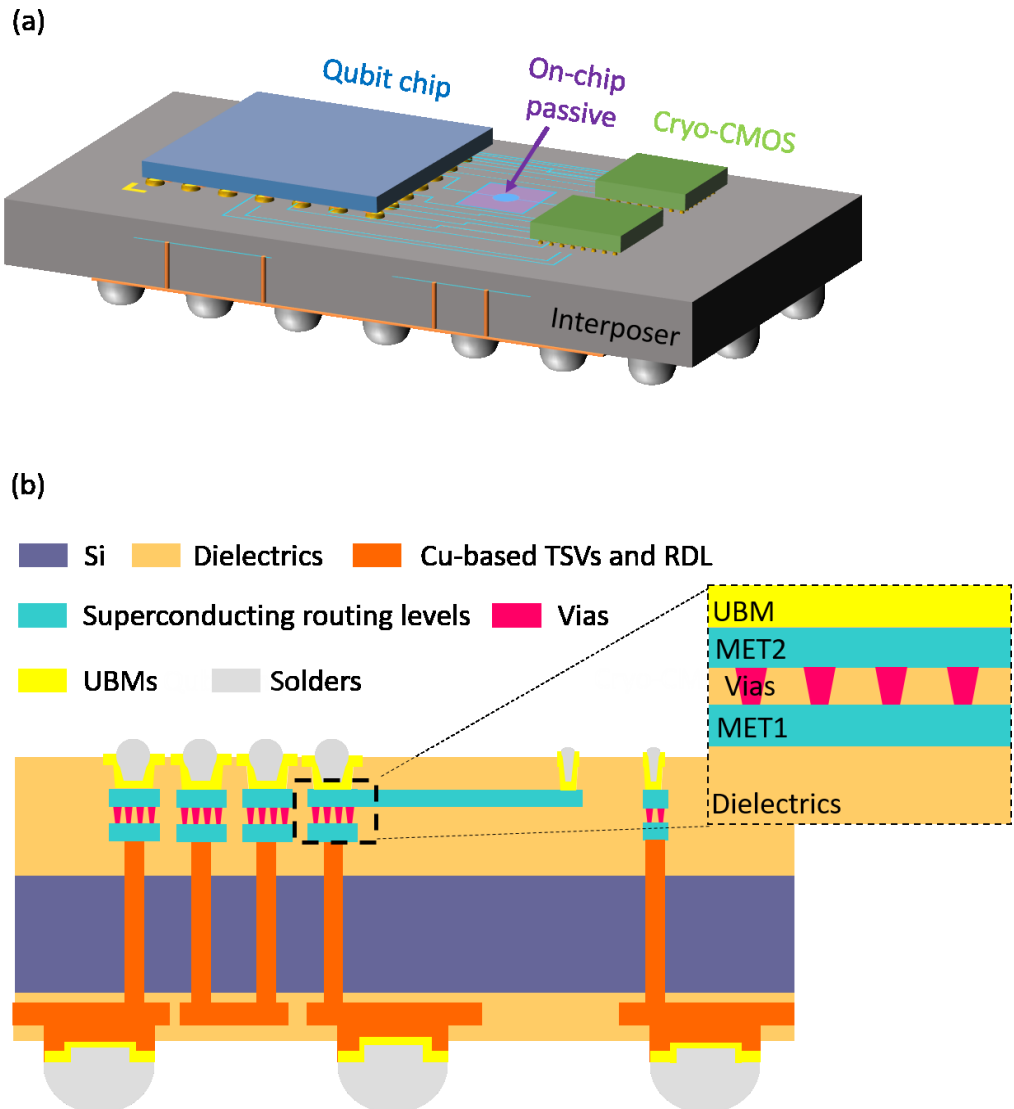

Figure 1- (a) Schematic representation of QuIC$^3$ demonstrator with one qubit chip and two cryo-CMOS ones on top of a Si-based interposer (see main text for acronym definition). (b) Focus on the interposer and its different technological levels. The black dotted square highlights the area of study in Figure 2 micrograph.

## 3. Interposer fabrication

The process flow starts with the plasma-enhanced chemical vapor deposition (PECVD) of SiO$_2$ dielectric on top of an oxidized 200 mm silicon wafer. The routing levels comprise a first layer, labeled as MET1, and a second layer, denoted MET2, which are connected through vias, as schematically illustrated in Figure 1(b). For the samples studied here, MET1 is fabricated with the magnetron sputtering of a Ti seed layer followed by a TiN barrier layer and a main $Al_{0.995}Cu_{0.005}$ layer, which is covered with Ti and TiN layers. This metallic stack, standard in microelectronics, is 540 nm thick. A small amount of Cu (here 0.5 %) is incorporated in the Al layer to improve its electromigration properties [31]. In the following, this stack will be referred as Ti/TiN/AlCu. MET2 is formed from either Ti/TiN/AlCu or a 200 nm-thick TiN, Nb or NbN layer. These deposits are performed in a multi-chamber plasma vapor deposition (PVD) tool with a substrate temperature of 400°C. MET1 and MET2 layers then experienced a standard photolithography and a plasma-assisted dry etch to define micrometer-wide tracks and components. A $SiO_2$ dielectric layer is then deposited on top of each metallic layer and planarized using chemical mechanical polishing. The vias connecting MET1 and MET2 are formed with a damascene process including, in this order, the deposition of $SiO_2$ dielectric, photolithography and etch steps to define openings, which are then filled with a Ti seed layer, a TiN barrier layer and W. Note that, for this study, a standard metallic stack was used to fill the vias but efforts are ongoing to fabricate fully superconducting vias by e.g. increasing the TiN thickness.

The routing levels are then covered with $SiO_2$ and SiN passivation layers, both deposited at 400°C. A photolithography step defines openings in the passivation stack to connect MET2 routing level to Au-plated UBM pads. Figure 2 displays a scanning electron micrograph evidencing all the front-side levels of the interposer, including MET1 and MET2 made here from Ti/TiN/AlCu and NbN, respectively. No defects are visible in the integration levels nor at their interfaces. Interposer front-side balling uses indium bumps, which are evaporated on top of the UBM as described in [32].

The complete process flow of the interposer front levels is schematically described in the Supplementary Information (S1). More details about the fabrication and morphological characterizations of MET1 and MET2 are also available in the Supplementary Information (S2).

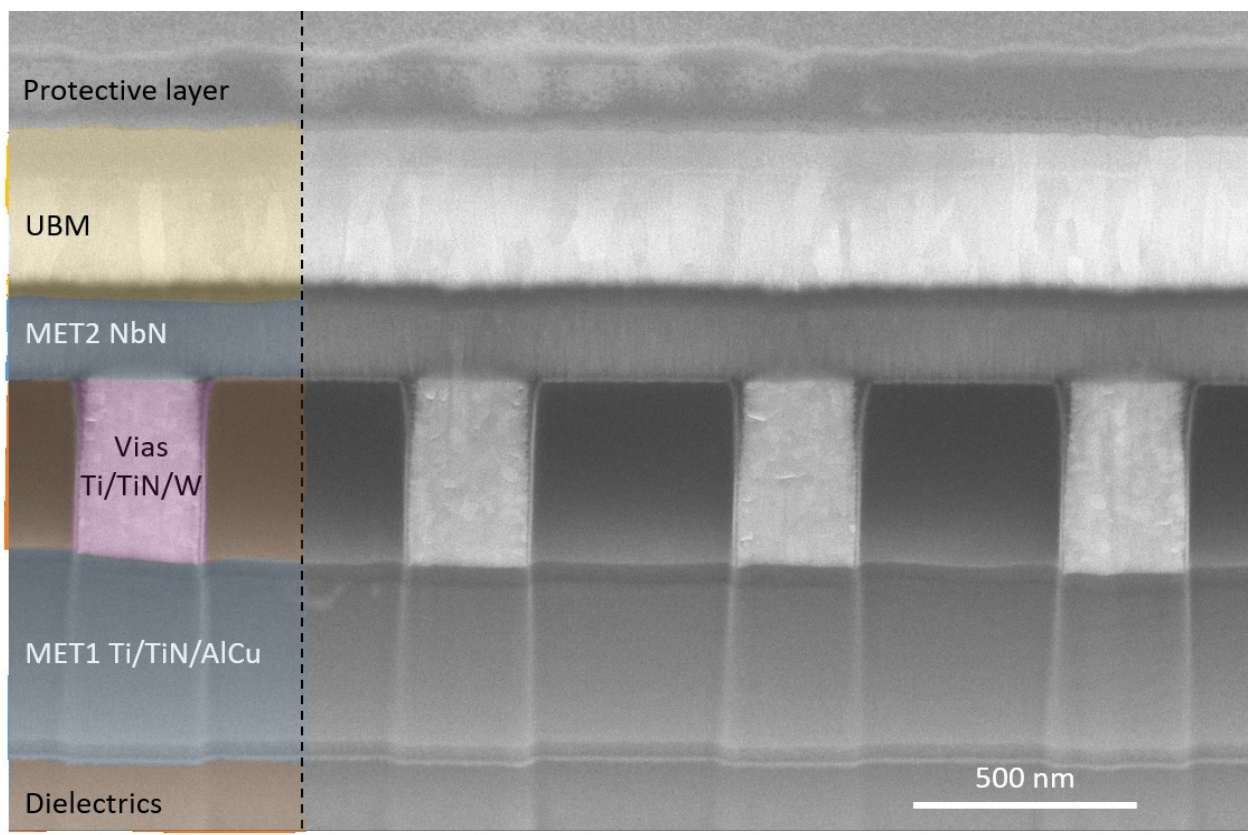

**Figure 2-** Focused ion beam scanning electron micrograph of the routing levels made from a MET1 layer in Ti/TiN/AlCu connected to a MET2 layer in NbN using Ti/TiN/W vias. The left-side of the micrograph has been false-colored to help identifying the layers.

## 4. Interposer electrical characterizations

After fabrication, the interposer wafers were tested both at 300 K on a fully automated probing station and at cryogenic temperature down to 2 K inside a cryostat. Tens of micrometer wide Van der Pauw devices and a few micrometer wide lines have been measured to extract the electrical and superconducting properties of the routing materials.

The results presented in the main text have been obtained on fully processed wafers with complete integration. Measurements have been performed, after the fabrication of the UBM, on devices integrated either on MET1 or on MET2 level. Additional data describing the impact of the integration steps on the electrical and superconducting properties of the routing layers are discussed in the Supplementary Information (S3a). Data acquired on chains connecting MET1 and MET2 levels are also presented in the Supplementary Information (S4).

*4.1 Parametric tests at 300 K*

Wafer-level parametric test results are presented in Figure 3 with the cumulative percentage as a function of the sheet resistance $R_{sq}$. This representation enables, in a single curve, to gain knowledge on the yield and dispersion of the measurements along the wafer. Measurements have been made on up to 45 fields distributed on the 200 mm wafer. The data are acquired on 2 μm-wide 500 μm-long lines of Ti/TiN/AlCu, TiN, Nb and NbN integrated on MET2 level. One can first notice that the cumulative percentage reaches 90 to 100% on all the presented measurements. The slight gap to 100%, observed for some materials, is essentially explained by non-complete field or fabrication defects on the dies located at the wafer edges. The mean values of the sheet resistances are indicated in Figure 3 for the different superconducting layers. With 71.0 mΩ/sq, Ti/TiN/AlCu routing is the less resistive compared to Nb, TiN and NbN, which are characterized by sheet resistances of 1.4 Ω/sq 3.3 Ω/sq and 21.2 Ω/sq, respectively.

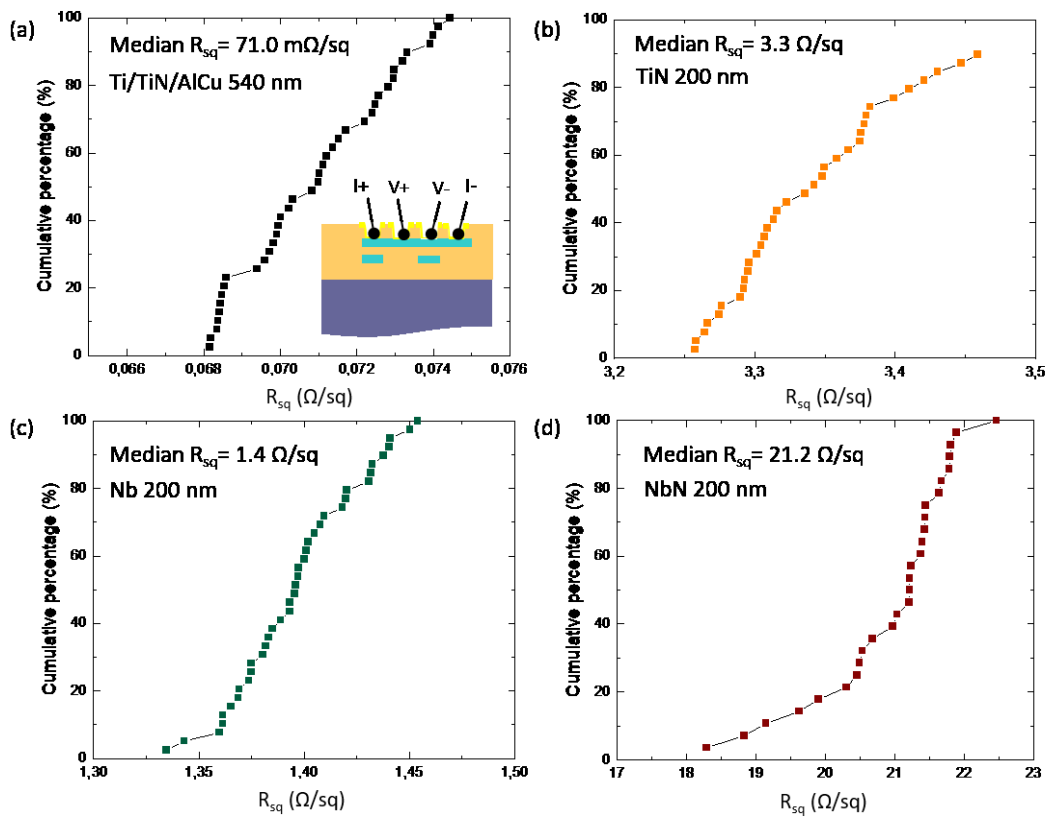

Figure 3-Cumulative % as a function of the sheet resistance $R_{sq}$ for (a) Ti/TiN/AlCu, (b) TiN, (c) Nb and (d) NbN MET2 routing levels. The measurements have been performed at 300 K on up to 45 fields along a 200 mm wafer after the full integration process, meaning after the UBM fabrication. Inset of (a) represents the measurement configuration using the color code of Figure 1(b).

*4.2 Resistance evolution with temperature and critical temperature*

Low frequency electrical measurements have been performed in a Physical Property Measurement System cryostat [33] with a base temperature of 2 K. The probed devices are 2 µm-wide channels with a length larger or equal to 500 µm, selected at the center of the wafers. Contrary to the parametric tests presented above, these measurements are done on single dies, which are cut and wire-bonded. Four-probe measurements associated with standard lock-in techniques are used to extract the resistances with a current bias of 4 µA. Figure 4 displays $R_{sq}$ as a function of the temperature $T$, between 300 K and 2 K, for all the investigated routing technologies. From these measurements, one can extract two properties: the residual resistivity ratio ($RRR$) and the critical temperature $T_c$. Defined in the following as $\frac{R_{sq}(300\,K)}{R_{sq}(T_c+1\,K)}$, the $RRR$ is a metrics of metal purity and defectivity [34]. Indeed, while at 300 K the resistance value is mostly driven by crystal vibration through electron-phonon interaction [35], this mechanism becomes negligible at low temperatures and the measured resistance mostly reflects the intrinsic imperfection levels of the probed metals. In Figure 4 (b), one can see the superconducting transitions of Ti/TiN/AlCu, TiN, Nb and NbN lines characterized by a vanishing of the resistances. The critical temperature $T_c$ is extracted at the center of the slope between the normal and the superconducting states. Table 1 summarizes the $T_c$ values, the $R_{sq}$ values at 300 K and at $T_c+1$ K, as well as the resulting $RRR$ for all the investigated routing layers.

For Ti/TiN/AlCu routing, Figure 4 presents measurements on lines integrated on either MET1 or MET2 levels. One can notice a few % difference on the $R_{sq}$ values with 71.7 mΩ/sq on MET2 and 76.3 mΩ/sq on MET1 at 300 K. This difference of resistance likely explains by the different thermal budgets experienced by MET1 and MET2 layers. Specifically, the 400°C passivation steps performed after MET1 and MET2 patterning lead to the formation of $Al_3Ti$ intermetallic known to slightly increase the resistance of this metal stack [36]. This intermetallic also introduces disorder as highlighted by the lower $RRR$ of MET1 (8.4) compared to MET2 (10.1). Note that additional data are presented in the Supplementary Information (S3a) evidencing the impact of the passivation step on the Ti/TiN/AlCu layer resistance. This also reflects on the critical temperature values with 3.9 K for MET2 and 3.7 K for MET1. Interestingly, $T_c$ values on both MET1 and MET2 layers are much larger than the 1.2 K expected for Al-based layers [35]. The proximity effect with TiN barriers elevates the superconducting properties of this stack [37][38]. Based on this result, in the following, Ti/TiN/AlCu will be characterized using the same procedures than the other tested type II superconductors. Concerning the TiN routing on MET2 layer, a critical temperature of 4.6 K and a $RRR$ of 1.3 are extracted. These values directly compare to a recent study of Faley *et al.* in which the integration of 100 nm-thick TiN layers in superconducting quantum interference devices and nanobridges results in a $T_c$ of 4.86 K and a $RRR$ of about 2 [39]. A larger $T_c$ of 7.7 K and a $RRR$ of 2.3 are obtained with Nb routing as can be seen in Figure 4 and Table 1. However, as discussed more in details in the Supplementary Information (S3a), $T_c$ and $RRR$ values of Nb are significantly impacted by the integration steps. A critical temperature of 9.3 K and a $RRR$ of 9.5 were indeed measured after patterning of the MET2 level. These latter values are slightly higher than the critical temperature of 9 K and the $RRR$ of about 6.5 recently reported by Verjauw *et al.* for the fabrication of Nb resonators on Si without any passivation on top or further integration steps [40]. Note that technological solutions (not discussed here) are currently investigated to limit the degradation of Nb superconducting properties during the integration. The largest $T_c$ is obtained for NbN routing with 14.0 K. This value is standard in the literature but might be optimized further by, e.g., fine-tuning $N_2$ flow rate [27]. This disordered superconductor is, from far, the most resistive routing technology investigated in this study and the only one to increase its resistivity when decreasing the temperature with 25.4 Ω/sq at 300 K and 35.6 Ω/sq at 15 K.

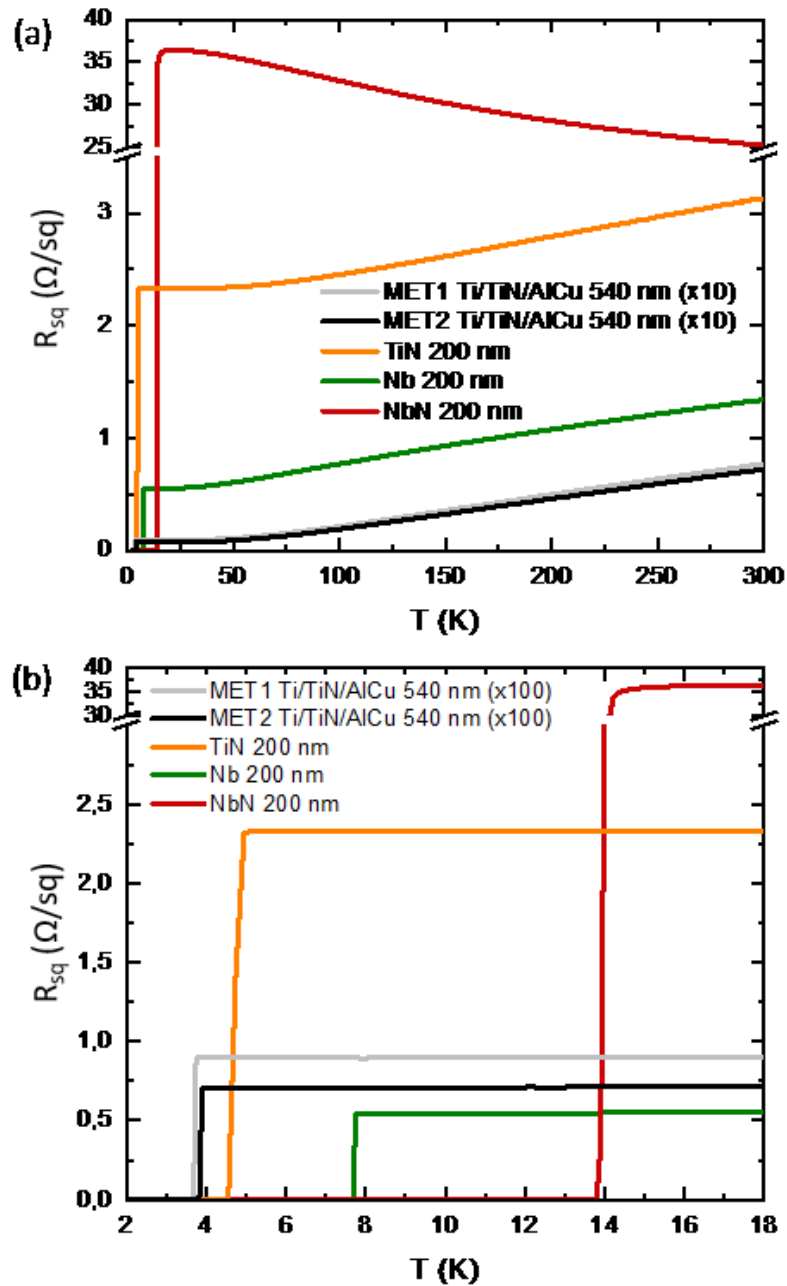

**Figure 4- (a)** Sheet resistance $R_{sq}$ evolution with the temperature *T* between 300 K and 2 K for Ti/TiN/AlCu routing on both MET1 and MET2 levels, and for TiN, Nb and NbN routing on MET2 level. Note that, in this plot, the resistance of Ti/TiN/AlCu devices has been increased by a factor 10 for clarity purpose. **(b)** $R_{sq}$ as a function of T between 18 K and 2 K for the different routing materials. Note that, in this plot, the resistance of Ti/TiN/AlCu devices has been increased by a factor 100 for clarity purpose.

*4.3 Critical magnetic field and critical current*

To complete the study of the routing layers, the critical magnetic field $B_{c2}$, characterizing the magnetic field driven superconducting to normal phrase transition in type II superconductors, and the critical current $I_c$ were also extracted, using a static magnetic field *B* perpendicular to the sample surfaces. Similar measurements with in-plane magnetic field were not possible here due to the cryostat set-up. Nevertheless, as in-plane critical field is typically larger than the out-of-plane one [41], the measured values of $B_{c2}$ give a low limit for the in-plane critical fields of Ti/TiN/AlCu, TiN, Nb and NbN integrated layers. Magnetic field values up to 8 T were needed to determine $B_{c2}$ and magnetic fields as low as 0.01 T and 0.1 T were applied for the critical current measurements. Note that a field of 0.1 T enabled to mimic the operating conditions of electric dipole spin resonance (EDSR) [42][43][44][45] and electron spin resonance (ESR) [46][47][48], commonly used to encode the information for spin qubits. The evaluation of the critical fields and currents is essential to optimize the design and sizing of the superconducting routing tracks and passive components on the interposer, as well as the required distance between those elements and the qubits.

Critical field measurements were performed on 2 µm-wide channels characterized with a length larger or equal to 500 µm. Figure 5 (a) displays $B_{c2}$ values extrapolated at 0 K as a function of $T_c$ for each investigated superconducting material. $B_{c2}$ was determined by repeating the resistance measurements as a function of the temperature for several values of magnetic field and by extracting the associated critical temperatures. The data and linear extrapolation used to extract $B_{c2}$ at 0 K are shown in the Supplementary Information (S3b) for TiN, Nb, NbN and Ti/TiN/AlCu channels. As summarized in Table 1, a minimum critical field of 2.1 T is obtained for Ti/TiN/AlCu routing on both MET1 and MET2 integration levels. A slight increase of $B_{c2}(0\ K)$ is noticed for Nb and TiN with 2.9 T and 3.2 T, respectively. NbN exhibits the larger critical field with 86.8 T. When considering these values, one should note that the fit used in this study does not take into account possible deviations from the linear relations at low temperatures [49].

The extraction of $B_{c2}(0\ K)$ enables to determine the coherence length $\xi(0\ K)$ for the superconducting routing layers using the following formula [50]:

$$\xi(0\ K) = \sqrt{\frac{\Phi_0}{2\pi B_{c2}(0\ K)}}$$

where $\Phi_0$ is the flux quantum. Table 1 displays $\xi(0\ K)$ values ranging from 1.9 nm for NbN to 12.5 nm for Ti/TiN/AlCu routing lines. Coherence length data are essential for the analysis of the critical current measurements as they dictate the line width $w$ at which magnetic vortices are likely to penetrate, thus affecting the superconducting properties. Specifically, for $w > 4.44 \times \xi(0\ K)$, vortex depinning is expected to limit the critical current, while for narrower channels ($w < 4.44 \times \xi(0\ K)$), the measured critical current is only attributed to depairing of Cooper pairs [51][52].

In this study, the critical current $I_c$ was evaluated on 2 µm-wide Ti/TiN/AlCu and Nb channels, a 6 µm-wide TiN line and a 25 µm-wide NbN one using I-V measurements with an excitation current of a few mA. These measurements were repeated for several temperatures close to $T_c$ and for different magnetic fields, 0 T, 0.01 T and 0.1 T, perpendicular to the samples. The widths of the devices have been chosen to minimize the resistance values in the order of hundreds of Ohm, limiting self-heating of the lines when transitionning to the normal state during the measurements. With such dimensions, presence of vortices is expected within the channels. With similar conditions, Engel *et al.* [52] have reported two distinct behaviors for the critical current:

(1) for temperature close to $T_c$, the critical current $I_c$ is similar to the depairing critical current $I_c^{GL}$ and can be described by the Ginzburg-Landau (GL) formula:

$$I_c^{GL}(t) = I_c^{GL}(0)(1-t^2)^{3/2}(1+t^2)^{1/2}, \text{ with } t = \frac{T}{T_c}$$

(2) for lower temperatures, the measured critical current is about half of the depairing critical current extracted in vortex-free channels.

Examples of I-V curves and associated fits using GL model are presented in the Supplementary Information (S3c) for TiN, Nb, NbN and Ti/TiN/AlCu routing options and for a perpendicular magnetic field $B$ up to 0.1 T. A rough estimation of the critical current value $I_c$ at low temperature is obtained from the approximation $I_c(0\ K) = I_c^{GL}(0\ K)/2$ based on [52]. The critical current density $J_c$ is then calculated using $J_c(0\ K) = \frac{I_C(0K)}{w \times d}$, with $d$ the thickness of the superconductor. The whole thickness of the Ti/TiN/AlCu stack was considered in this calculation as it is expected to be fully superconductive by proximity effect.

Figure 5 displays $J_c$ as a function of $B$ for the different superconducting routing. At zero magnetic field, Nb exhibits the larger critical current density with 9.9 MA/cm². This value agrees with the study of Huebener *et al.* [53] in which a 1 µm-thick Nb film, deposited in the same conditions at 400°C, had a critical current density estimated to about 7 MA/cm². This also compares to Il'In *et al.* [54] measurements on nanometer-thick Nb films where $J_c$ was probed in between 5 and 30 MA/cm² for micrometer wide devices. The critical current density of TiN is slightly lower with 2.0 MA/cm². It is an order of magnitude lower for NbN with 0.6 MA/cm² and two orders of magnitude lower for Ti/TiN/AlCu with 0.06 MA/cm².

The large effect of magnetic field is evidenced by a significant decrease of $J_c$ for all materials. For example between 0 T and 0.1 T, Ti/TiN/AlCu critical current density has dropped from 85% down to 0.009 MA/cm². The reduction is similar for Nb with 86% variation down to 1.3 MA/cm² and even higher for TiN with a 95% decrease of $J_c$ down to 0.1 MA/cm². Impact of the magnetic field is less pronounced for NbN with only a 50% reduction of the critical current down to 0.3 MA/cm² at 0.1 T. This agrees with the high magnetic field resilience of NbN highlighted here by its large perpendicular critical magnetic field and reported in literature through the preservation of its quality factor with a few hundreds of mT perpendicular magnetic field [28].

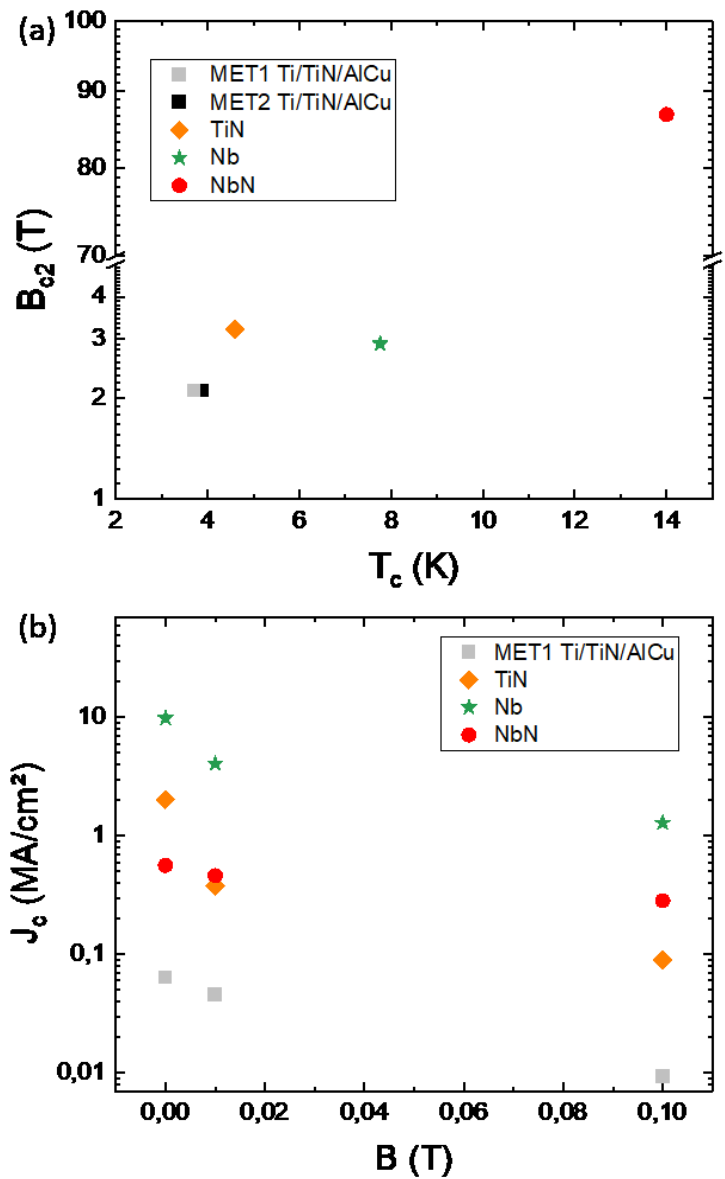

**Figure 5-** Superconducting properties of TiN, Nb, NbN and Ti/TiN/AlCu routing lines. (a) Critical perpendicular magnetic field $B_{c2}$ *(0 K)* vs critical temperature $T_c$. (b) Evolution of the critical current density $J_c$ *(0 K)* as a function of the perpendicular magnetic field *B*.

Table 1- Electrical and superconducting properties of 540 nm-thick Ti/TiN/AlCu and 200 nm-thick TiN, Nb and NbN routing layers.

| Materials | Integration level | $R_{sq}$ at 300 K ($\Omega$/sq) | $R_{sq}$ at $T_c$ + 1 K ($\Omega$/sq) | RRR | $T_c$ (K) | $B_{c2}$ (0 K) (T) | $\xi$(0 K) (nm) | $J_c$ (0 K) (MA/cm²) at B=0T | $J_c$ (0 K) (MA/cm²) at B=0.1T |
|---|---|---|---|---|---|---|---|---|---|
| Ti/TiN/AlCu | MET2 | $7.17 \times 10^{-2}$ | $0.71 \times 10^{-2}$ | 10.1 | 3.9 | 2.1 | 12.5 | Not measured | Not measured |
| Ti/TiN/AlCu | MET1 | $7.63 \times 10^{-2}$ | $0.89 \times 10^{-2}$ | 8.4 | 3.7 | 2.1 | 12.5 | $63.5 \times 10^{-3}$ | $9.4 \times 10^{-3}$ |
| TiN | MET2 | 3.13 | 2.32 | 1.3 | 4.6 | 3.2 | 10.1 | 2.0 | 0.1 |
| Nb | MET2 | 1.35 | 0.54 | 2.3 | 7.7 | 2.9 | 10.7 | 9.9 | 1.3 |
| NbN | MET2 | 25.40 | 35.64 | 0.7 | 14.0 | 86.8 | 1.9 | 0.6 | 0.3 |

*4.4 Discussion*

Table 1 summarizes the electrical and superconducting properties of the Ti/TiN/AlCu, TiN, Nb and NbN investigated routing layers. With critical temperatures above 3.7 K, all these materials will be in the superconducting phase at the operating temperature of Si spin qubits, in between a few mK to 1.5 K [2]. However, to minimize the electronic thermal transport, to e.g. thermally decouple the qubits and their control electronics, the materials with larger $T_c$ such as Nb and NbN will be preferred. Indeed, the larger the ratio between $T_c$ and $T$ is, the less unpaired electrons remains to participate to the thermal transport.

When performing spin manipulation, static magnetic field up to a few hundreds of mT can be used with EDSR [43][44][45] or ESR [47][48] techniques. With critical fields $B_{c2}$ larger than 2.1 T and up to 86.8 T for NbN, the integration of the investigated superconductors appears possible to form routing tracks and passive components dedicated to read-out. Preliminary RF characterizations of embedded passive components will be discussed in the next section.

Several types of signals are flowing through the interposer with different current ranges. One can consider the input and output currents of qudot charge measurements, which are typically in the order of a few pA to a few nA [47]. Larger currents, in the order of a few hundreds of μA up to a few mA, are required for cryo-CMOS circuit supplies and along metal ESR striplines to generate an oscillating magnetic field used to manipulate the spin [46]. To limit thermal dissipation within the interposer and more specifically prevent heating from Joule effect, the supply lines carrying such current ranges should optimally be superconducting. At very low temperatures, the estimated critical current densities of the tested superconductors are all compatible with the design of routing tracks and ESR lines as narrow as a few tens of nm to carry currents larger than 1 mA at magnetic field up to 0.1 T. This is also true when the operating temperature is increased around 1 K as the critical current densities only experience a few % decrease at this temperature as can be seen in Figure 17(b), Figure 18(b), Figure 19(b) and Figure 20(b) of the Supplementary Information. In these conditions (*T*=1 K, *B*=0.1 T), tracks larger than 100 nm would nevertheless be preferred for Ti/TiN/AlCu to carry mA current while preserving its superconducting phase. It is worth noting that for each materials, the main constraint to the design rules appears to be the lithography capabilities rather than the material properties. Typically, electron beam lithography enables to fabricate channels as narrow as a few tens of nm with the exact width depending on several parameters such as the equipment, the material and resist thicknesses as well as the used exposure. By decreasing *w* in the tens of nm range, it might become comparable to *4.44×ξ(0 K)*, which may have an impact on the superconducting properties.

Based on all these results, there is no clear counter-indication to the integration of Ti/TiN/AlCu, TiN, Nb and NbN as superconducting routing layers in the interposer or even as part of the back end of line of the qubit and cryo-CMOS circuits. The co-integration of these superconductors on the same routing levels or on successive ones can be wise to optimize the whole system. For example, with its large critical current density, Nb routing could be used on MET1 to supply μA to mA currents to the cryo-CMOS circuits while NbN, which is the best candidate to reduce thermal transport, could connect the cryo-CMOS and qubit chips through the interposer MET2 level. The superconducting properties of Ti/TiN/AlCu and TiN are less attractive than those of Nb and NbN but the compatibility and maturity of these materials with actual industrial processes motivate their integration.

## 5. Integration of passive RF devices

Various RF passive components, such as resistors, inductors and capacitors, have been designed and embedded on the interposer routing levels to give insights on the routing and passive properties up to a few GHz, which is the frequency range typically used to encode and read spin qubits. The devices were characterized at cryogenic temperature using a Lakeshore EMPX-HF cryogenic probe station and the frequency-dependent S-parameters were measured from 10 MHz to 26.5 GHz with an Agilent N5222A network analyzer. Two ground-signal-ground probes were used and short-open-load-through calibration was performed at low temperature using a calibration substrate [55]. Open and short structures were measured to estimate parasitic effects. Contact resistance between probes and landing pads was found to have the most significant impact with values ranging from minus to plus 0.3 Ω at low frequencies, affecting the measurement of superconducting devices. Within these conditions, no de-embedding procedure was achieved. During the probe station cool-down, the superconducting transition of NbN at 14 K was observed while the one of Nb expected at 7.7 K was not visible. Based on these observations and the thermal dissipation between the probes and the sample holder, the sample temperature was estimated to be around 10 K.

Figure 6 displays typical measured frequency dependent characteristics of three different passive components: a resistor, an inductor and a capacitor. These devices have been conceived using MET1 routing level in Ti/TiN/AlCu and MET2 in NbN connected by Ti/TiN/W vias. The resistor was made from a chain with 700 links, each link comprising 2 µm-wide MET1 and MET2 sections connected by 2x2 arrays of vias. Its short-circuit resistance exhibits a stable behavior up to a few GHz with a value of 356 Ω at 1 GHz (see Figure 6(a)). This value agrees with the low frequency resistance measured at 10 K on a similar resistor (see Figure 22 of Supplementary Information). Figure 6 (b) demonstrates the functionality of a 14.5 turn spiral inductor (labeled Ind A in the following) formed with 2 µm-wide superconducting traces on MET2. The serial inductance equals 22 nH at 1 GHz with a resonance frequency larger than 10 GHz. A metal-insulator-metal 50 µm-wide square capacitor formed through the $SiO_2$ dielectric layer between MET1 and MET2 routing levels reaches 200 fF over a wide frequency range as shown in Figure 6 (c).

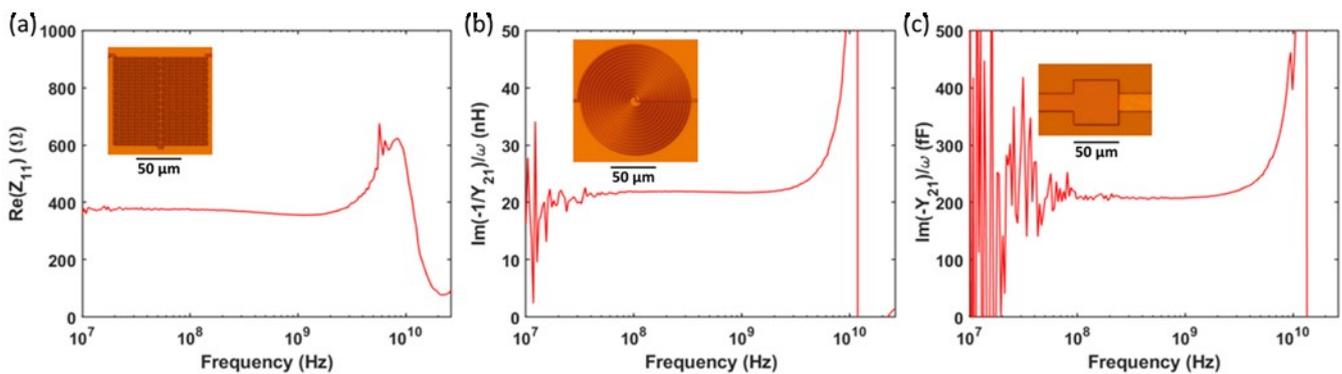

**Figure 6- Measured frequency dependence of a (a) resistor (b) inductor and (c) capacitor fabricated with NbN as MET2. Insets show optical micrographs of the components. The temperature of the components is estimated to 10 K during the measurements.**

Focusing on inductors, single layer devices made with trace widths down to 1 µm on MET2 enable to reach a very broad range of inductance values using small footprints. Table 2 describes three tested inductors, Ind A, Ind B and Ind C for which the number of turns, the inner and outer diameters as well as the trace and space widths have been varied [56], targeting inductances ranging from 10 to 500 nH. Optical micrographs of these inductors are available in the Supplementary Information (see Figure 23). The short-circuit inductances as a function of frequency are presented in Figure 7 for these three inductors. Note that the same inductors have been fabricated using either Ti/TiN/AlCu as MET2 which is not superconducting at the measurement temperature or NbN which, on the other hand, is in its superconducting phase. The inductors made with Ti/TiN/AlCu as MET2 perform as expected by the models with values of 11 nH, 99 nH and 491 nH at 100 MHz, corresponding to the calculated geometric inductances. Inductors made with superconducting NbN as MET2 exhibit higher values of inductances with 21.8 nH, 172 nH and 820 nH at 100 MHz. This difference is the result of the kinetic inductance of the disordered NbN superconductor [28], which is estimated between 7 and 10 pH/sq in these measurements.

**Table 2 - Geometric characteristics of the inductors and associated measured inductances.**

| Devices | Number of turns | Outer diameter (µm) | Trace space and width (µm) | Calculated geometric inductance (nH) | Measured inductance (nH) with Ti/TiN/AlCu at 100 MHz | Measured inductance (nH) with NbN at 100 MHz |
|---|---|---|---|---|---|---|
| Ind A | 14.5 | 122 | 2 | 10.5 | 11 | 22 |
| Ind B | 16.5 | 328 | 2 | 105 | 99 | 172 |
| Ind C | 59.5 | 286 | 1 | 505 | 491 | 820 |

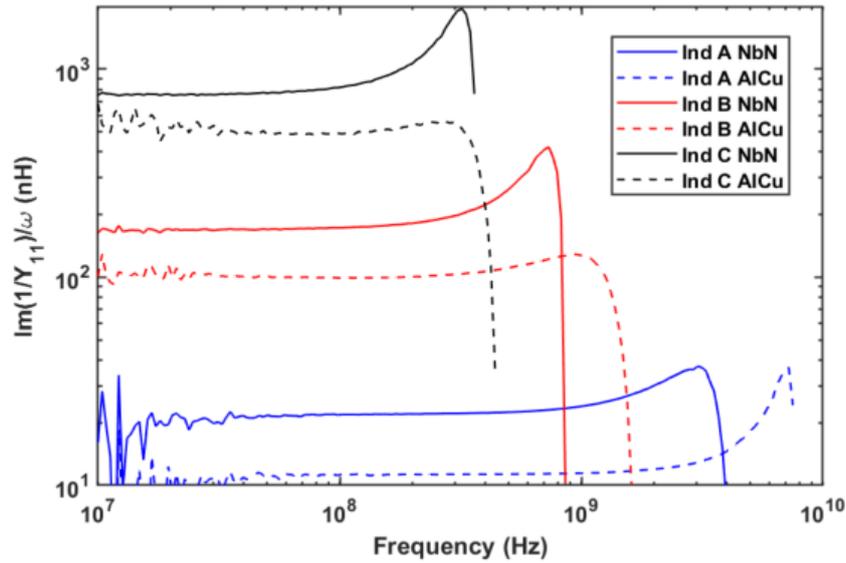

**Figure 7-** Measured frequency dependence of short-circuit inductance for three inductors geometries (Ind A, Ind B and Ind C) with NbN (straight lines) or Ti/TiN/AlCu (noted AlCu, dashed lines) as MET2.

For Ti/TiN/AlCu-based inductors, the resistance values extracted in the tens to hundreds of MHz range (see Figure 7) are in agreement with the resistivity determined in the low frequency tests (see Figure *3*4), with 19 Ω, 62 Ω and 300 Ω for Ind A, Ind B and Ind C, respectively. For the superconducting NbN-based inductors, the extracted series resistances below hundreds of MHz compare to the measurement uncertainties, making complex the extraction of the quality factor. Nevertheless, these preliminary measurements suggest that super-inductors with impedances exceeding the resistance quantum [57] and quality factors higher than 100 are obtained.

## 6. Conclusion

Using 200 mm Si wafer technologies, superconducting routing levels were integrated on a Si-based interposer. Aiming to electrically couple spin qubit arrays with cryo-CMOS control and read-out circuits, this platform was characterized at low temperatures to extract its routing properties. Specifically, micrometer-wide lines made of Ti/TiN/AlCu, TiN, Nb and NbN were measured to identify their superconducting regime in terms of temperature, magnetic field and current. All these materials have shown their compatibility and adequate superconducting properties within Si spin qubit operating conditions, meaning temperature up to ~1 K and static perpendicular magnetic field up to a few hundreds of mT. Besides, the extracted critical current densities of TiN, Nb and NbN open the way toward the integration of lines as narrow as a few tens of nm to carry currents as large as a 1 mA at 1 K and 0.1 T. Hundreds of nm wide routing tracks will be preferred for Ti/TiN/AlCu in these extreme conditions. Additionally, preliminary low temperature RF measurements on resistors made with MET1 and MET2 levels in Ti/TiN/AlCu and NbN confirmed the stability of the routing up to GHz. More than evaluating the properties of the superconducting routing levels, the low temperature RF measurements of integrated devices such as resistors, capacitors and inductors highlight the promises of superconducting RF passive elements, showing extensive impedance values over a wide frequency range. These results motivate the integration of superconducting routing levels in interposers or even in the back end of line of spin qubit and cryo-CMOS circuits to optimize the whole system performances. While the superconducting properties of Ti/TiN/AlCu and TiN are inferior to those of Nb and NbN, the industrial compatibility of these materials justify their consideration in integration schemes. This work is expected to also benefit to superconducting qubit integration and to other fields requiring superconductor integration such as astrophysics [58].

**Acknowledgements**

This work was partially funded by the European Research Council Grant QuCube and French government through Nano2022. The authors would like to thank the CEA Si 200 mm platform, Gaël Pillonnet, Simon Guignard and Joris Lacord for their contribution and support to the circuit design, Rémi Vélard for the FIB imaging, Roselyne Templier for the AFM study, Daniel Braithwaite for the access to the PPMS cryostat, Laurent Vila, Cécile Grès, Aurélie Kandazoglou and Quentin Berlingard for their support with the RF cryoprober.

**Supplementary information**

### S1) Interposer process flow

The process flow of the interposer front-side fabrication is schematically described in Figure 8.

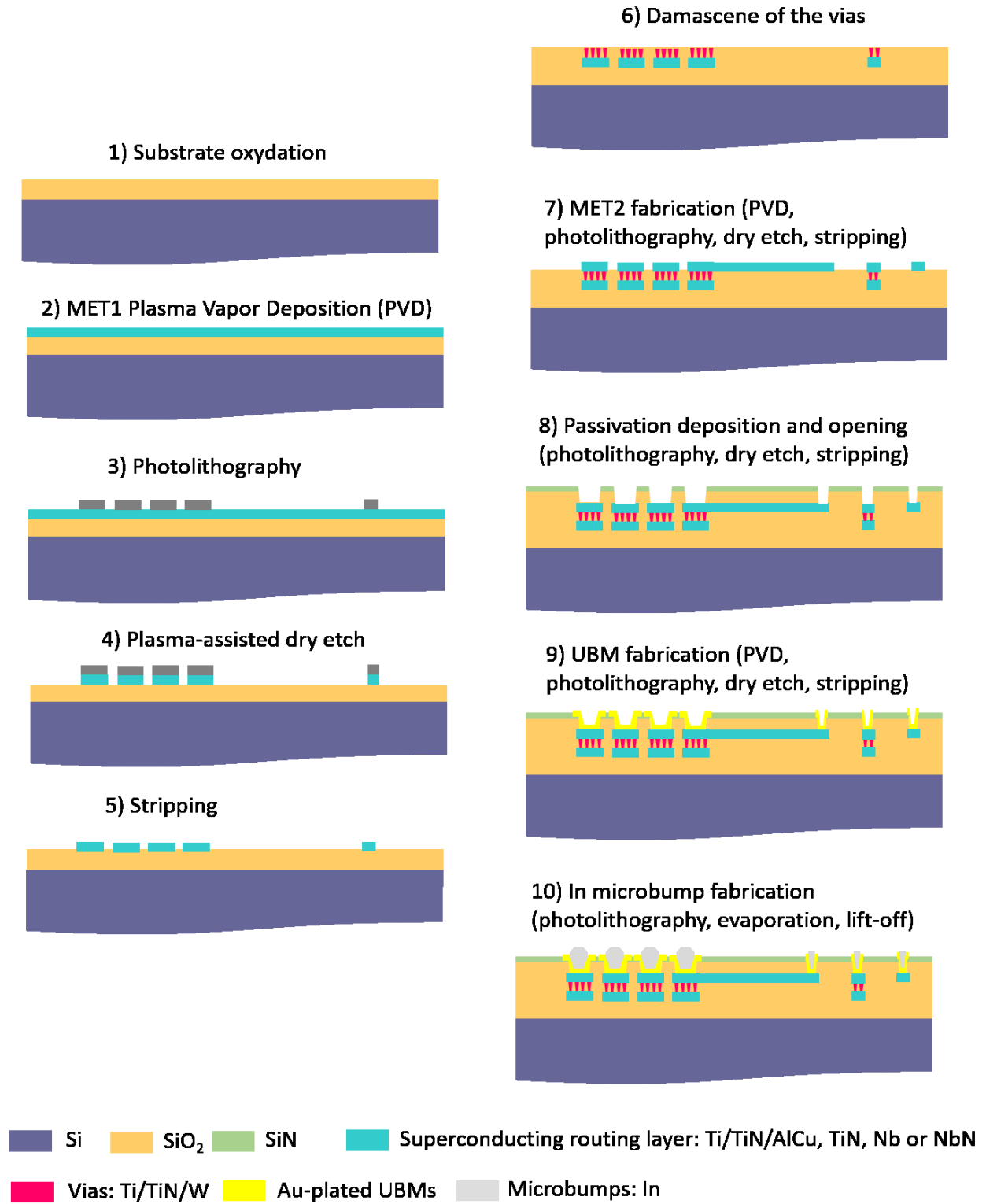

**Figure 8- Schematic representation of the complete front-side process flow of the interposer.**

**S2) Fabrication and morphological characterizations of the Nb and NbN routing levels**

Different materials and stack layers have been used in this study to form MET1 and MET2 superconducting layers. MET1 layer has mostly been made from a standard Ti/TiN/AlCu stack, while MET2 layer can integrate the same stack or more attractive superconductors such as TiN, Nb or NbN. While the fabrication of Ti/TiN/AlCu and TiN tracks use mature recipes developed for years for microelectronics, engineering studies were needed to develop proper etching conditions for Nb and NbN with $CF_4$ as primary ion etching gas. Figure 9 and Figure 10 display morphological characterizations of Nb and NbN tracks, respectively, using best-developed etch and stripping processes so far. A slight over-etching (tens of nm) of the $SiO_2$ template, that is yet to optimize, is observed for both cases on Figure 9(e) and Figure 10(e). Nonetheless, the roughness of the $SiO_2$ and the superconductor (< 3nm) are not impacted by the etching and stripping steps.

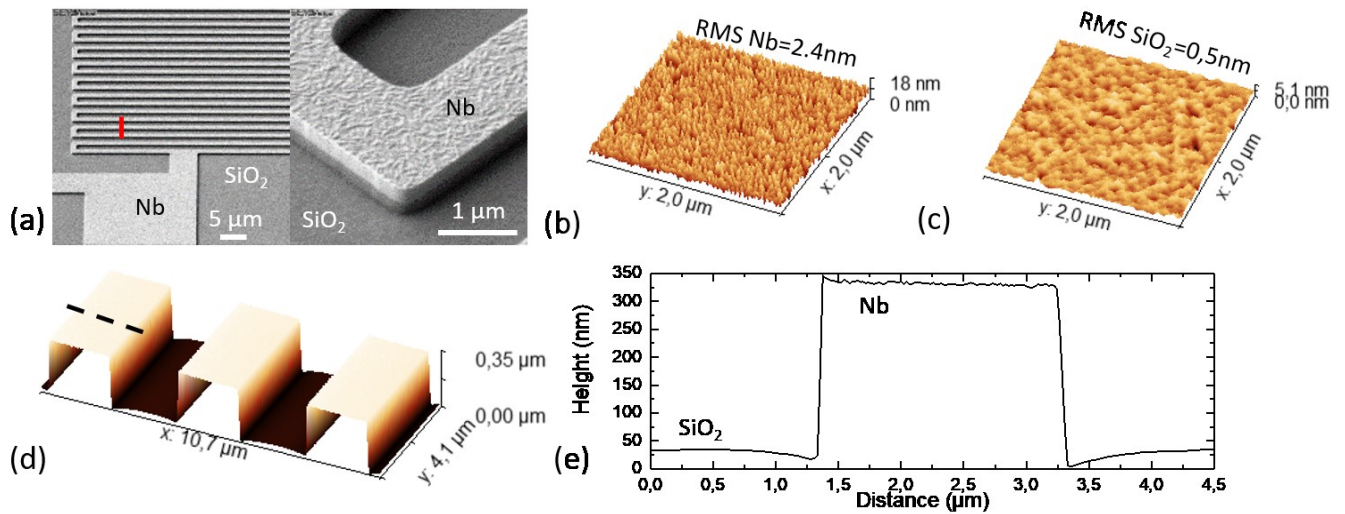

**Figure 9-** Morphological characterizations of Nb routing tracks. **(a)** Scanning electron micrographs of the Nb tracks just after etching. **(b)** and **(c)** 2x2 µm² atomic force micrographs on the Nb and $SiO_2$, respectively. **(d)** 3D map of the tracks of (a), taken along the red line. **(e)** 1D profile extracted along the black dashed line of (d).

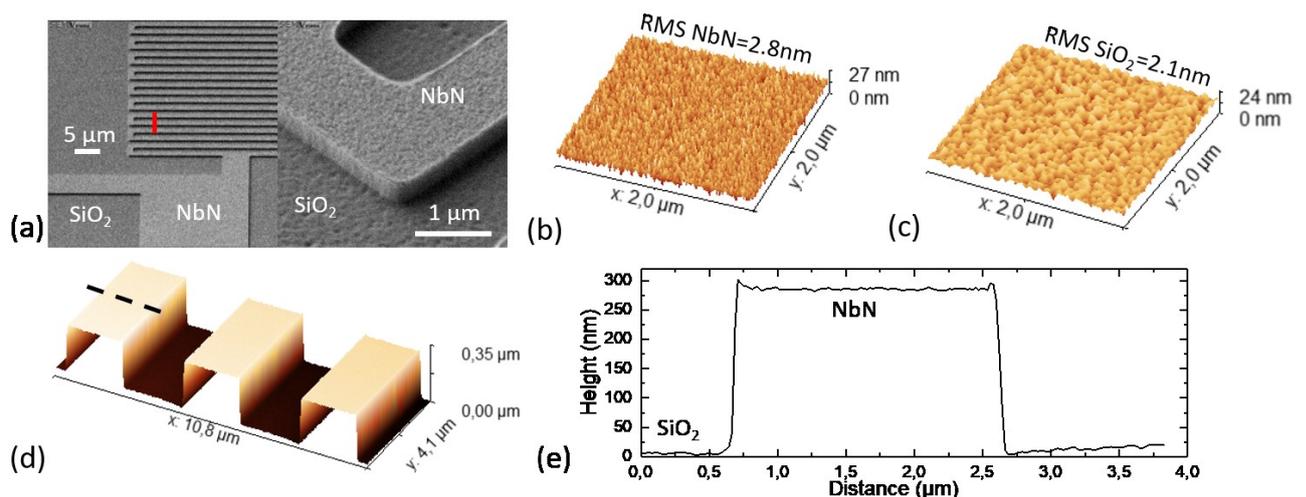

**Figure 10-** Morphological characterizations of NbN routing tracks. **(a)** Scanning electron micrographs of the NbN tracks just after etching. **(b)** and **(c)** 2x2 µm² atomic force micrographs on the NbN and $SiO_2$, respectively. **(d)** 3D map of the tracks of (a), taken along the red line. **(e)** 1D profile extracted along the black dashed line of (d).

### S3) Electrical characterizations of superconducting routing levels

#### S3a) Impact of the integration steps on *Rsq*, *T$_c$* and *RRR*

Wafer-level parametric tests have been performed at 300 K along the process to evaluate the impact of each integration steps (MET2 patterning, passivation deposition and UBM fabrication) on the routing layer electrical properties. Figure 11 displays the cumulative percentage of the sheet resistances along the wafers for Ti/TiN/AlCu, TiN, Nb and NbN layers. TiN integration appears robust with no resistance variation along the process and a sheet resistance value of 3.35 +/- 0.10 Ω/sq (see Figure 11 (b)). Concerning Ti/TiN/AlCu routing, a slight increase of about 3% of the sheet resistance, from 0.069 +/- 0.003 Ω/sq to 0.071 +/- 0.003 Ω/sq, is noticed after the passivation of the metal. This is attributed to the formation of $Al_3Ti$ intermetallic when the substrate reaches 400°C for the passivation deposition [36]. The resistance is not impacted further by the UBM fabrication for which the thermal budget is much lower.

Similar behavior is observed for NbN on Figure 11(d) with a 7% increase of the resistance from 20.0 +/- 0.7 Ω/sq to 21.5 +/- 0.8 Ω/sq after the passivation and UBM fabrication. Effects of the integration steps are mostly observed in the case of Nb as can be seen in Figure 11 (c). An increase of about 30 %, from 1.07 +/- 0.04 Ω/sq to 1.55 +/- 0.06 Ω/sq, of the resistance is measured after passivating the metal. The insulating oxide $Nb_2O_5$ is likely to form at the interface between Nb and $SiO_2$ increasing the resistance of the lines [59]. The fabrication of the UBM is accompanied by a reduction of the resistance (~ -10%) down to 1.39 +/- 0.06 Ω/sq. A more detailed study, including TEM imaging, would be needed to properly understand the modification of Nb film and its properties during the integration.

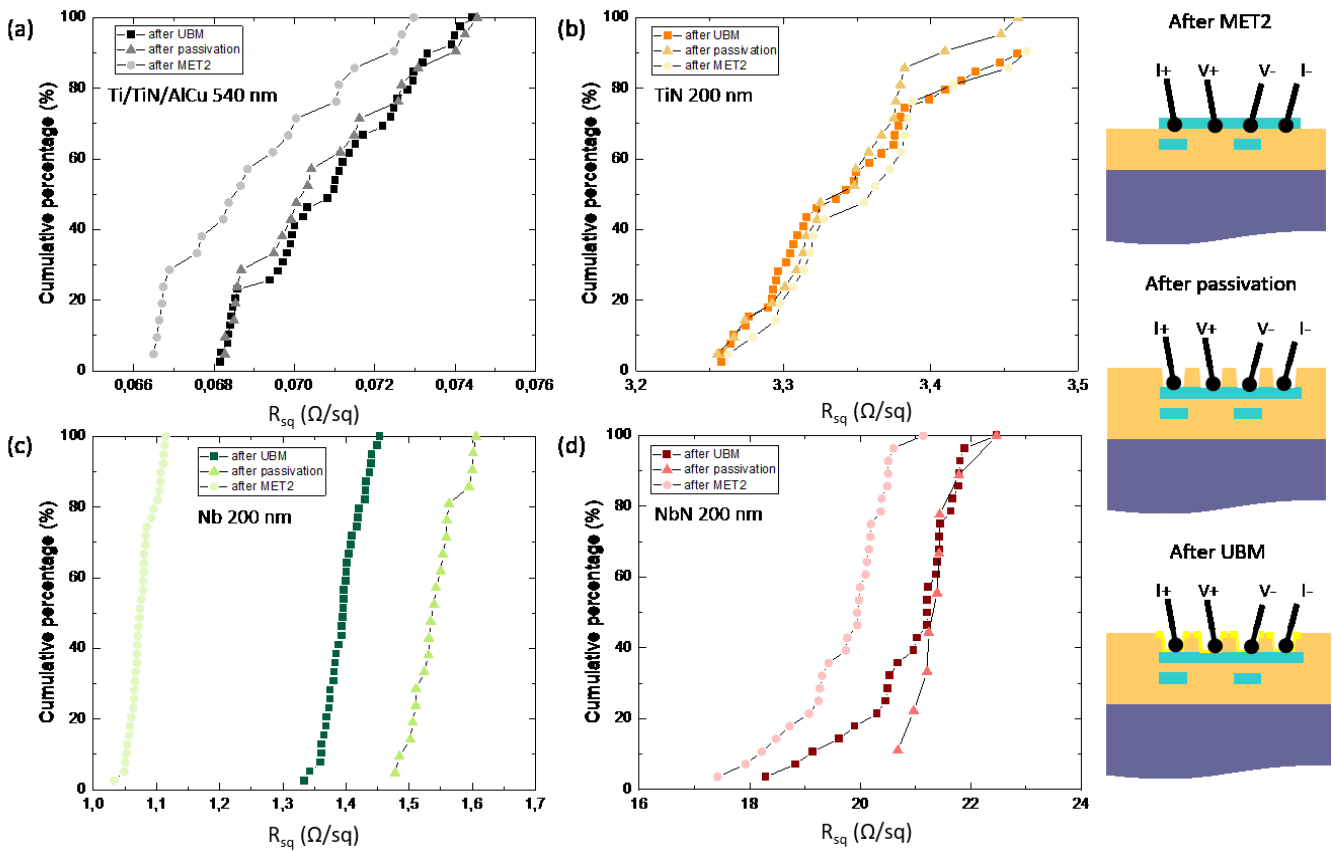

**Figure 11- 300 K parametric test results for (a) Ti/TiN/AlCu, (b) TiN, (c) Nb and (d) NbN routing layers. Each panel shows data obtained after MET2 integration step, after passivation deposition and after the UBM fabrication.**

Figure 12 presents the evolution of *R$_{sq}$* as a function of the temperature between 300 K and 2 K for TiN, Nb and NbN routing at the different integration steps. As shown on Figure 12 (a) and (c), passivation and UBM integraton steps have no impact on the resistance evolution with temperature and so on the *RRR* for TiN and NbN. The critical temperature is also unchanged. The situation is different for Nb. At 300 K, consistently with the parametric tests discussed just before, one can already see an increase of $R_{sq}$ after passivation followed by a decrease of 10% after UBM step. The difference of $R_{sq}$ at low temperatures are also significant leading to the following *RRR* variations: 9.5 after MET2 patterning, 2.3 after depositing the passivation and 2.5 after fabricating the UBM. The reduction of the *RRR* by a factor of about 3 and the increase of $R_{sq}$ values after passivation suggest an increase of the disorder in the Nb routing layer. This could be attributed to a significant oxidation during the process and by the interdiffusion with the $SiO_2$ layer on top. The critical temperature is also affected with a reduction from 9.3 K to 7.7 K. Technological solutions (not discussed here) are currently investigated to protect the Nb layers from oxidation and from intermixing with $SiO_2$ passivation layer during the integration, to optimize the properties of this routing option.

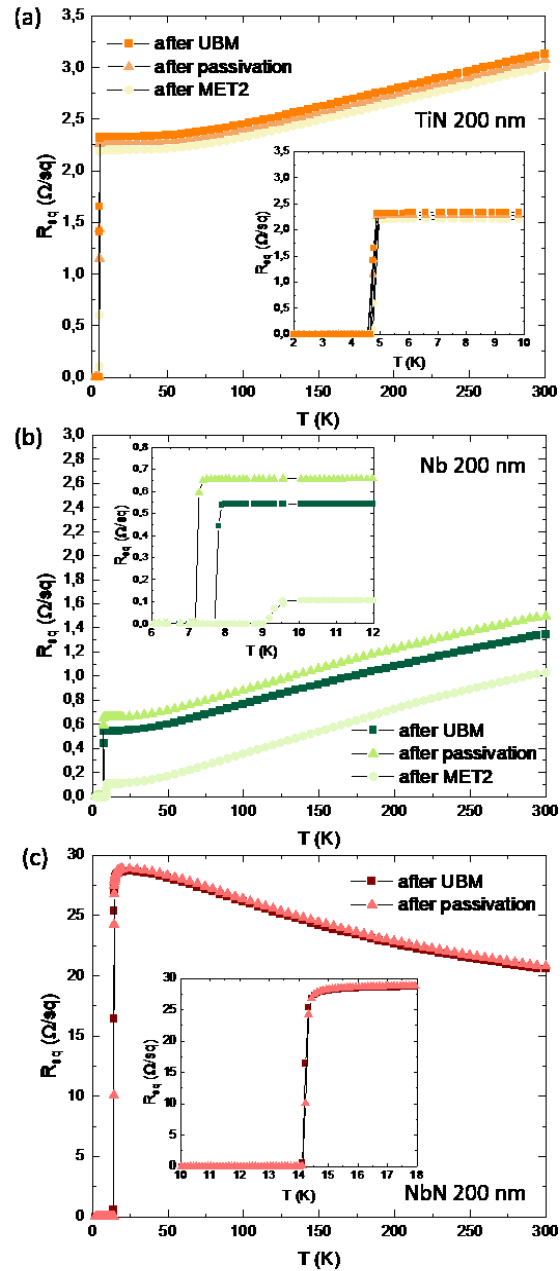

**Figure 12-** $R_{sq}$ **as a function of** *T* **between 300 K and 2 K for (a) TiN, (b) Nb and (c) NbN after the realization of MET2 level, after the passivation deposition and opening and after the UBM fabrication. Insets focus on** $R_{sq}$ **variations at low temperatures to highlight the superconducting transitions.**

### S3b)    Extraction of the critical magnetic field

Resistance measurements as a function of the temperature were repeated for several values of the static perpendicular magnetic field *B* for the different routing materials as displayed in Figure 13 for NbN, Figure 14 for Nb, Figure 15 for TiN and in Figure 16 for Ti/TiN/AlCu routing layers. The extraction of the critical temperature at each magnetic field is reported in the panels (b), showing a linear correlation between $T_c$ and *B*. The linear extrapolation of the data is used to estimate the critical field value $B_{c2}(0K)$ for each routing materials based on $\xi(T) = \frac{\xi(0)}{\sqrt{1-\frac{T}{T_C}}}$ and $B_{c2}(T) = \frac{\Phi_0}{2\pi\xi^2(T)}$ [49].

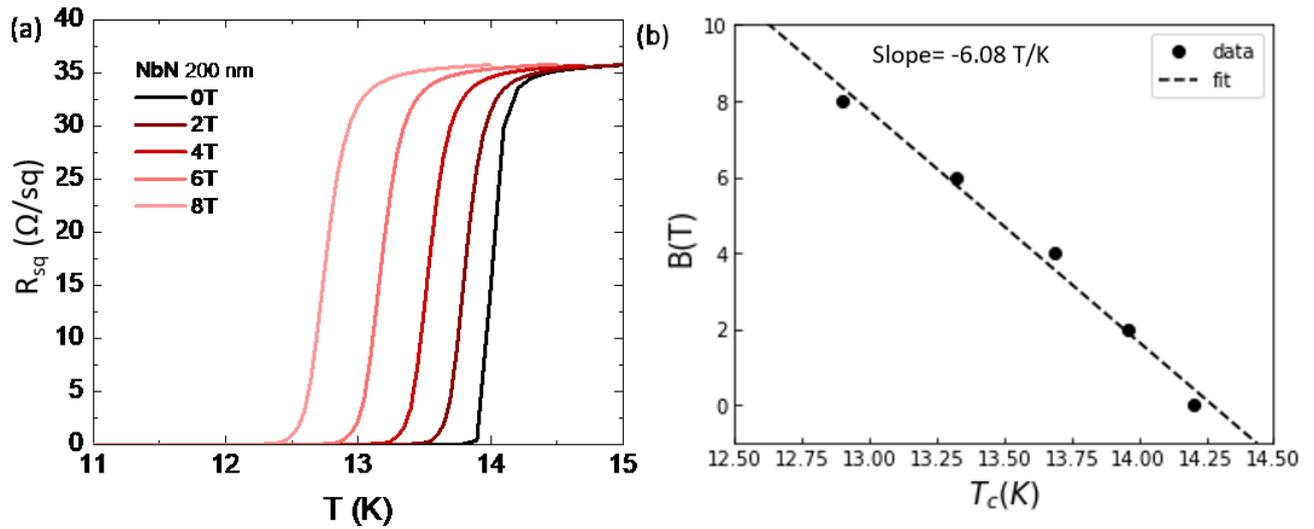

**Figure 13-(a)** Sheet resistance $R_{sq}$ as a function of the temperature for different magnetic field *B* values in between 0 T and 8 T for 200 nm-thick NbN routing layer. **(b)** Evolution of the critical temperature $T_c$ as a function of *B* and the associated linear fit.

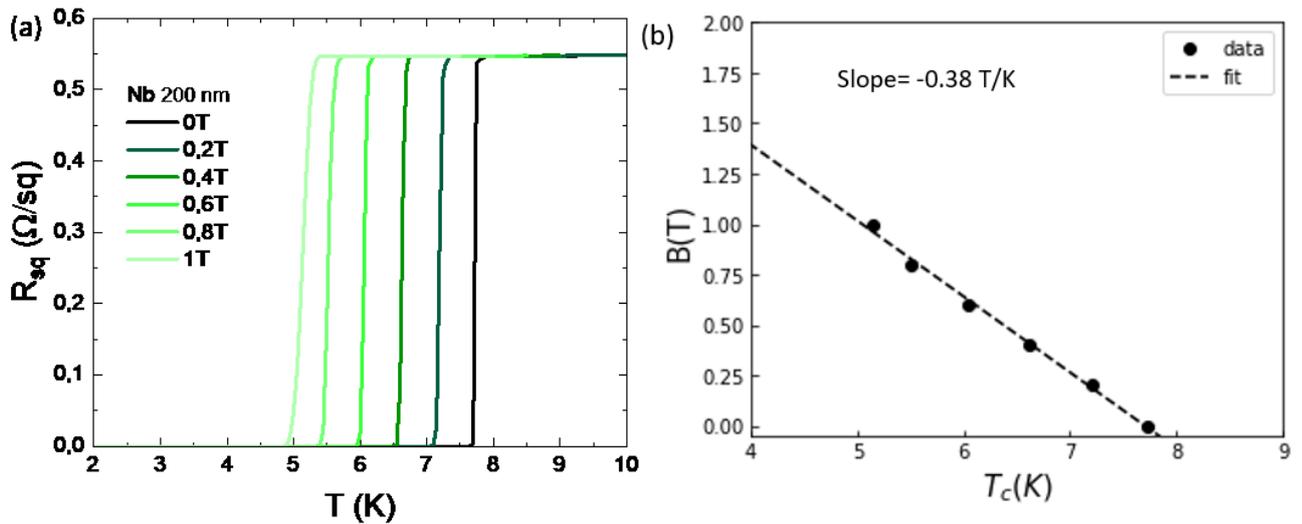

**Figure 14-(a)** Sheet resistance $R_{sq}$ as a function of the temperature for different magnetic field *B* values in between 0 T and 1 T for 200 nm-thick Nb routing layer. **(b)** Evolution of the critical temperature $T_c$ as a function of *B* and the associated linear fit.

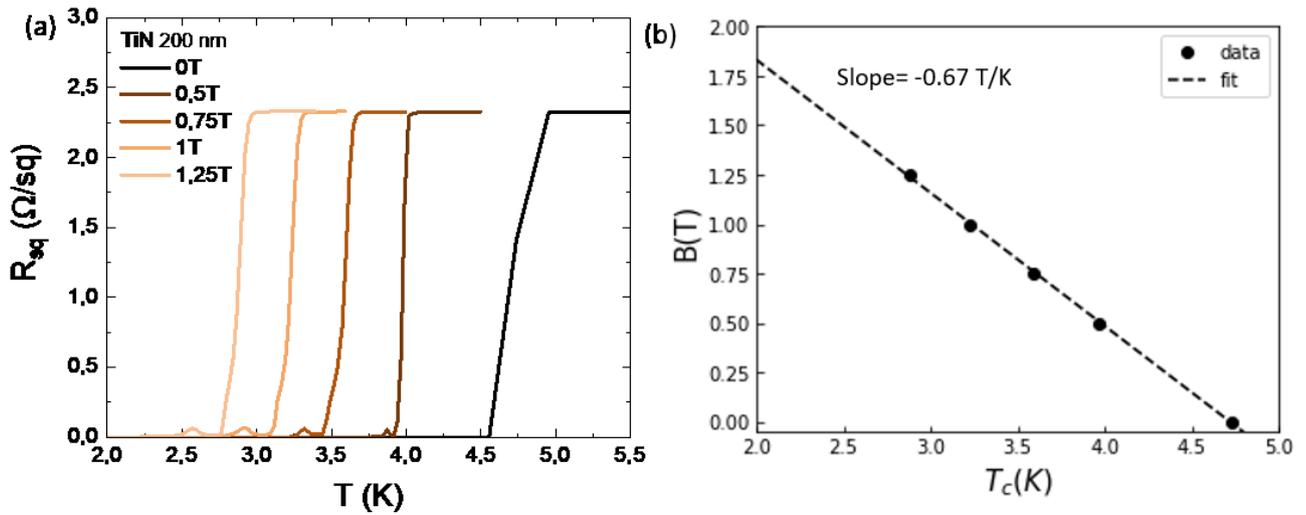

Figure 15-(a) Sheet resistance $R_{sq}$ as a function of the temperature for different magnetic field *B* values in between 0 T and 1.25 T for 200 nm-thick TiN routing layer. (b) Evolution of the critical temperature $T_c$ as a function of *B* and the associated linear fit.

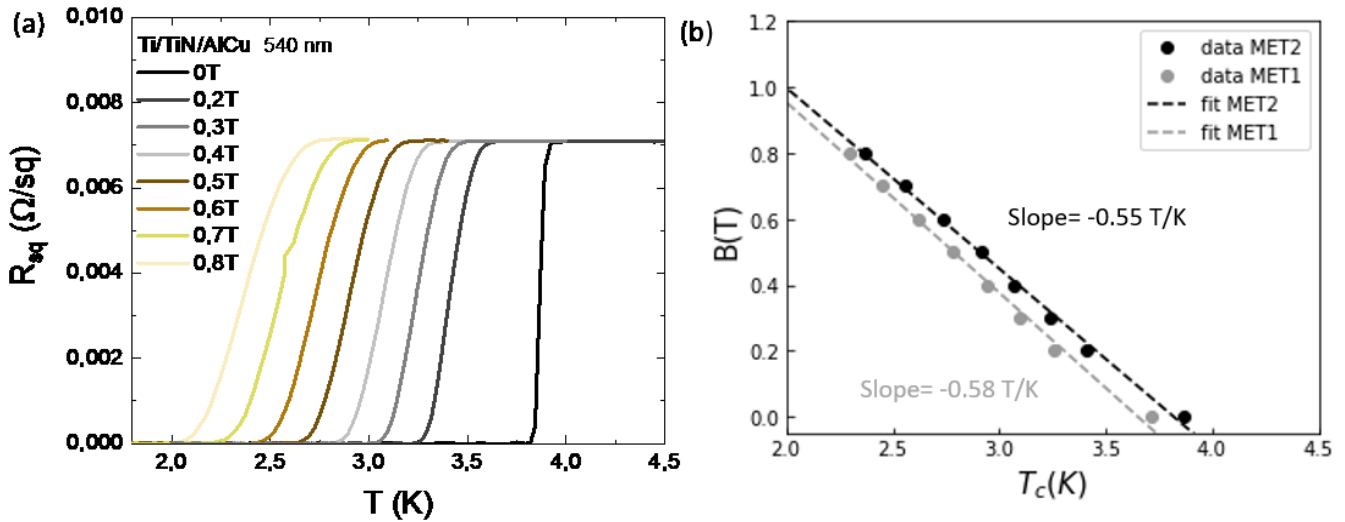

Figure 16-(a) Sheet resistance $R_{sq}$ as a function of the temperature for different magnetic field *B* values in between 0 T and 0.8 T for 540 nm-thick Ti/TiN/AlCu routing layer on MET2 integration level. (b) Evolution of the critical temperature $T_c$ as a function of *B* and the associated linear fit.

### S3c) Extraction of the critical current

Figure 17 (a), Figure 18 (a) and Figure 19 (a) present I-V measurements performed on a 25 µm wide NbN channel at T=13 K, on a 2 µm wide Nb channel at 6.5 K and on a 6 µm wide TiN channel at 3.5 K, respectively. The width of the channel was chosen to limit the resistance value in the order of hundreds of Ohm for these measurements. For each material, the excitation current $I_{ex}$ was swept from 0 mA to a positive value and then swept back to 0 mA (data not shown here) after reaching the superconductor-normal transition. The critical current $I_c$ value corresponds to the $I_{ex}$ value at which the voltage abruptly switches to a non-zero value. These measurements were done for several temperatures close to $T_c$ and for different magnetic fields, 0T, 0.01T and 0.1T, perpendicular to the sample. Note that the gain limits in the PPMS cryostat prevented from doing I-V measurements for which $I_{ex}$ exceeded ~20 mA, preventing measurements at $T/T_c \ll 1$. Each measurement was repeated four times with a very small dispersion of $I_c$ as denoted by the barely visible error bars in Figure 17 (b). $I_c$ experimental values are plotted as a function of $T/T_c$ in Figure 17(b), Figure 18(b) and Figure 19(b) for NbN, Nb and TiN channels, respectively. These experimental data are fitted using Ginzburg-Landau theory to estimate the critical current $I_c^{GL}(0\ K)$. As explained in the main text, we then consider $I_c(0\ K) = I_c^{GL}(0\ K)/2$ [52].

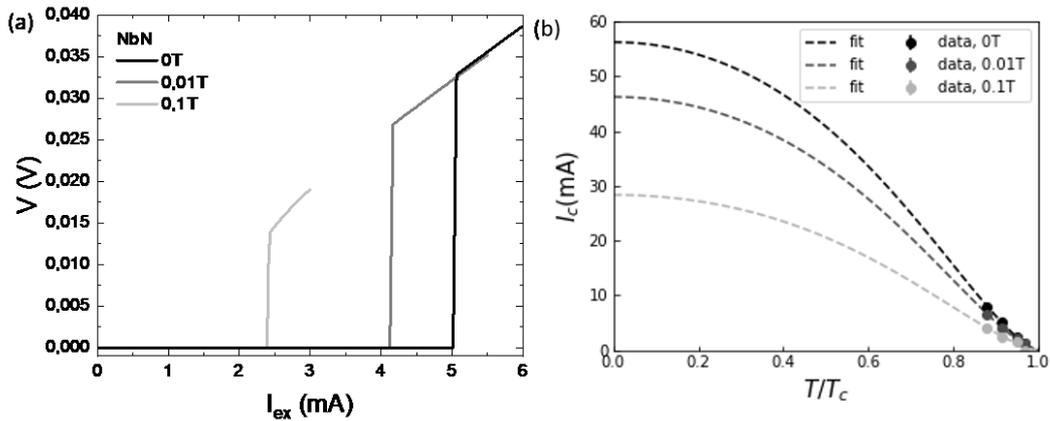

**Figure 17- (a)** V-I curves obtained at 0T, 0.01T and 0.1T at T=13 K for a 200 nm-thick NbN routing layer. **(b)** Evolution of the measured critical current *$I_c$* as a function of *$T/T_c$* at 0T, 0.01T and 0.1T and the associated fits of *$I_c^{GL}$* using Ginzburg-Landau model.

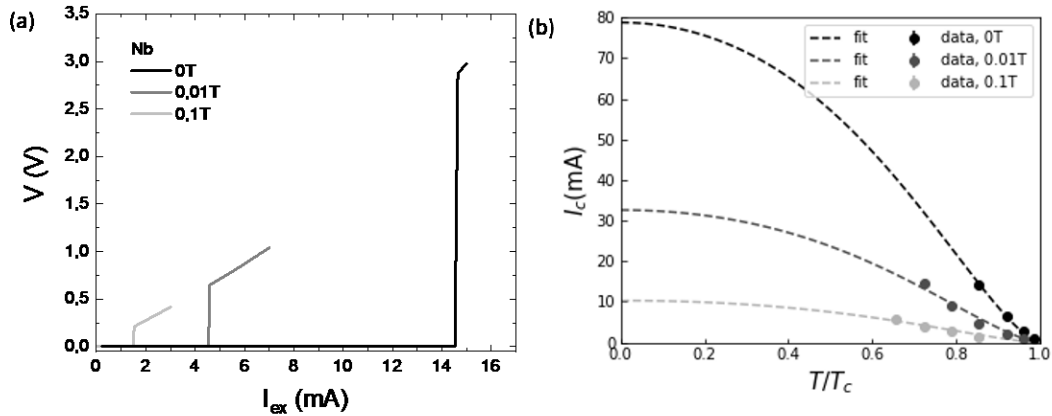

**Figure 18- (a)** V-I curves obtained at 0T, 0.01T and 0.1T at T=6.5 K for a 200 nm-thick Nb routing layer. **(b)** Evolution of the measured critical current *$I_c$* as a function of *$T/T_c$* at 0T, 0.01T and 0.1T and the associated fits of *$I_c^{GL}$* using Ginzburg-Landau model.

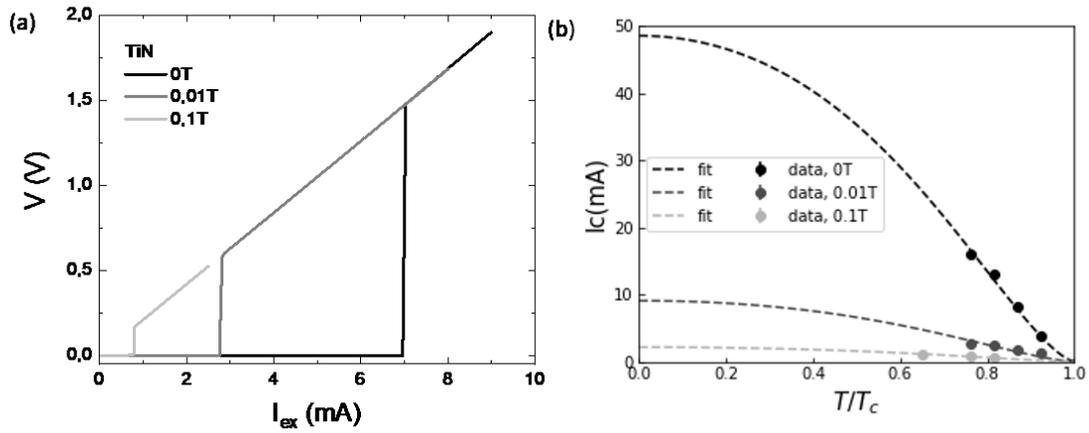

**Figure 19-** (a) V-I curves obtained at 0T, 0.01T and 0.1T at T=3.5 K for a 200 nm-thick TiN routing layer. (b) Evolution of the measured critical current $I_c$ as a function of $T/T_c$ at 0T, 0.01T and 0.1T and the associated fits of $I_c^{GL}$ using Ginzburg-Landau model.

Similar measurements were performed on Ti/TiN/AlCu as displayed in Figure 20. This time, the I-V curve shape does not show an abrupt transition from 0 to a positive voltage but a smooth increase of the voltage with $I_{ex}$. The presence of two superconductors TiN and AlCu in the stack layer might explain this behavior. To extract $I_c$ from these curves, a 1 µV criterion was used.

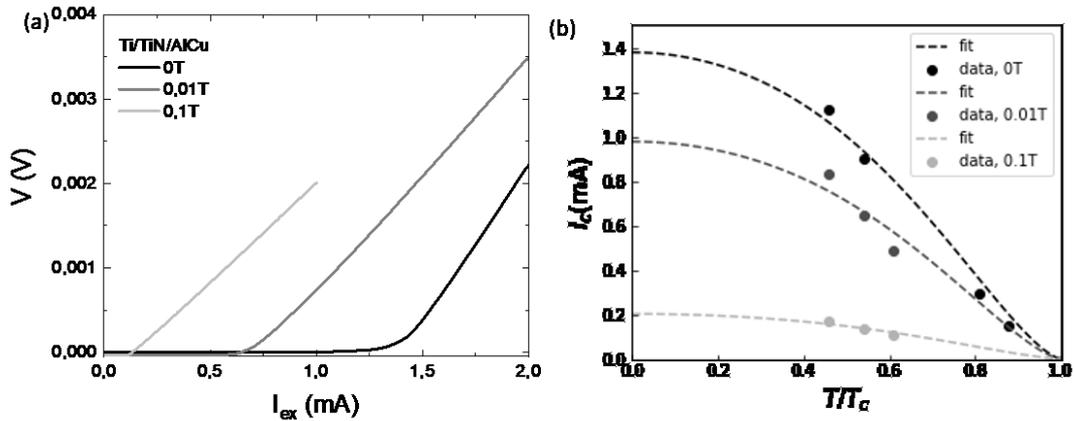

**Figure 20-** (a) V-I curves obtained at 0T, 0.01T and 0.1T at T=2 K for a 540 nm-thick Ti/TiN/AlCu routing layer integrated on MET1 level. (b) Evolution of the measured critical current $I_c$ as a function of $T/T_c$ at 0T, 0.01T and 0.1T and the associated fits of $I_c^{GL}$ using Ginzburg-Landau model.

### S4) Electrical characterizations of MET1-MET2 chain

The electrical characterizations presented in the main text focus on a single layer. Here we show additional data obtained on a chain connecting 2 µm-wide MET1 and MET2 sections. MET2 is made from Ti/TiN/AlCu, TiN, Nb or NbN layers and MET1 from Ti/TiN/AlCu. Four Ti/TiN/W vias are used at each connection between MET1 and MET2 layers. Figure 21 presents the link resistance values obtained on chains made with 7000 links between MET1 and MET2 layers. The link resistance is obtained by dividing the total resistance of the chain by the number of links. These parametric tests have been performed at 300 K on an automated probing station. One can first notice that the measurement yield is larger than 95% on the 200 mm wafers for all the probed chains, highlighting the proper electrical contact between MET1 layer, the vias and MET2 layer no matter what superconductor was integrated as MET2. The dispersion observed when MET2 is made of Ti/TiN/AlCu, TiN and Nb is mostly explained by slight resistance variations or defects at the wafer edges that affect either MET1 level, the vias or MET2 level, depending on the tested dies.

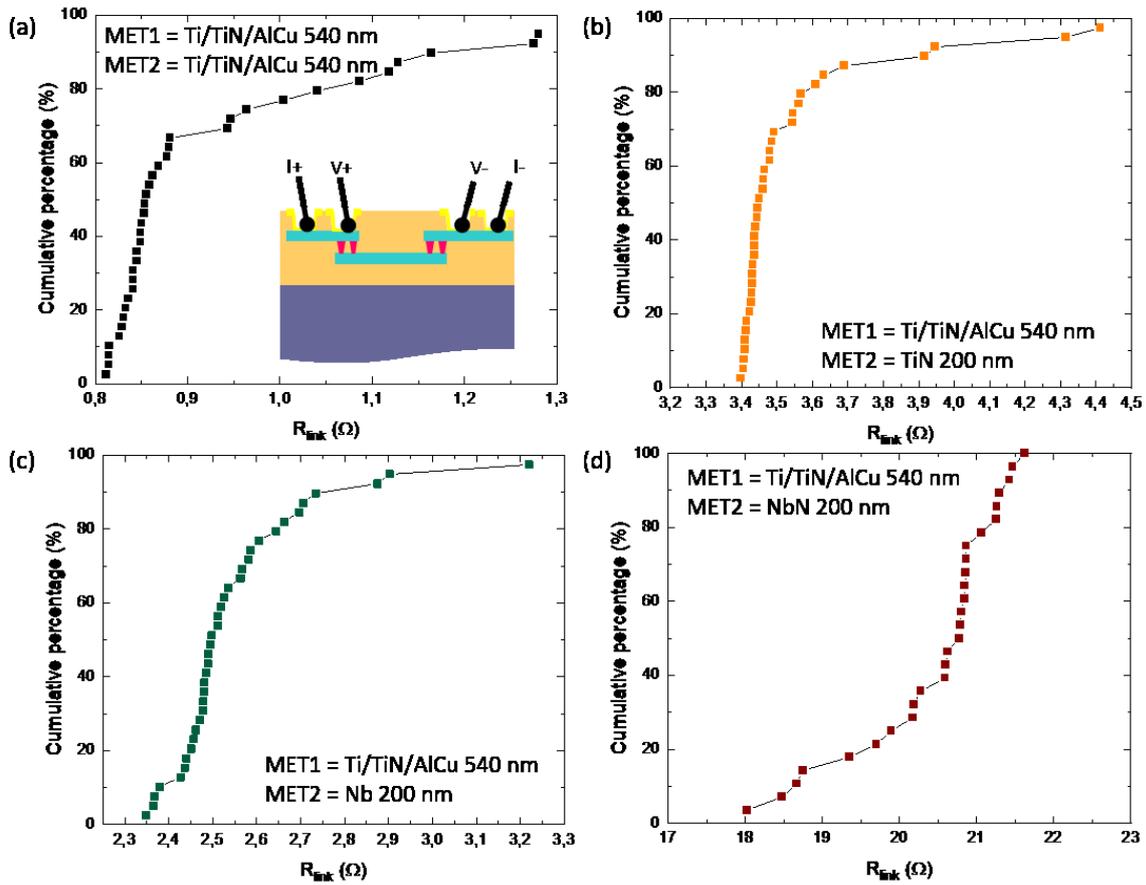

**Figure 21- Cumulative % as a function of the link resistance $R_{link}$ for chains made of MET1 layer in Ti/TiN/AlCu and MET2 layer in (a) Ti/TiN/AlCu, (b) TiN, (c) Nb and (d) NbN. Inset of (a) represents the measurement configuration using the color code of Figure 1(b).**

Complementary to these parametric tests, Figure 22 presents the evolution of the link resistance as a function of the temperature, between 300 K and 2 K, in the case where the chain includes NbN on MET2. In the normal state, one can see that the resistance varies similarly to a single layer of NbN (see Figure 12 (c)) due to the large difference (a factor of about 300) of $R_{sq}$ existing between NbN on MET2 and Ti/TiN/AlCu on MET1. At low temperatures, two transitions are observed (see the black arrows in the inset of Figure 22): a sharp one at 14 K corresponding to NbN superconducting-normal transition on MET2 and a smoother one at around 3.6 K attributed to Ti/TiN/AlCu transition on MET1. The sharpness of the transitions is explained by difference of sheet resistance existing between the chain blocks. For lower temperatures, the resistance value corresponds to the four Ti/TiN/W vias connecting the layers. It linearly decreases down to 2 K. Additional measurements at lower temperatures would be needed to fully characterize this chain at the operating temperature of spin qubits.

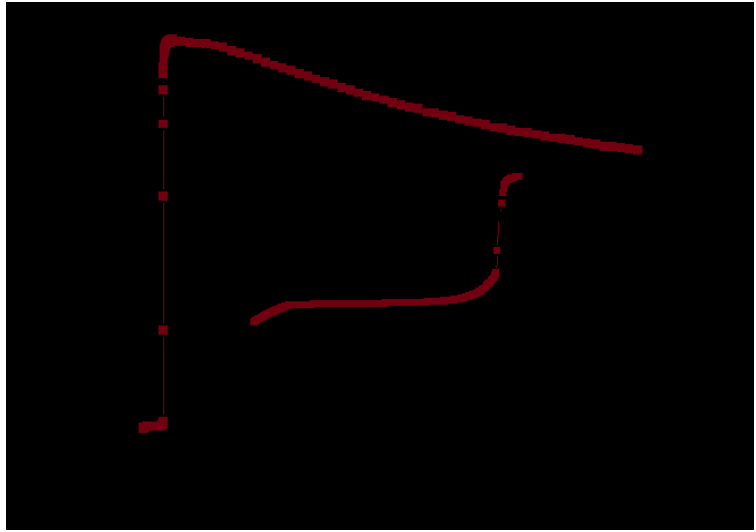

**Figure 22-Link resistance *R*<sub>link</sub> as a function of the temperature *T* for a chain made with MET1 in Ti/TiN/AlCu, Ti/TiN/W vias and MET2 in NbN. The inset shows a zoom at low temperatures.**

### S5) Optical micrographs of the inductors
The fabricated single layer inductors Ind A, Ind B and Ind C are visible in Figure 23.

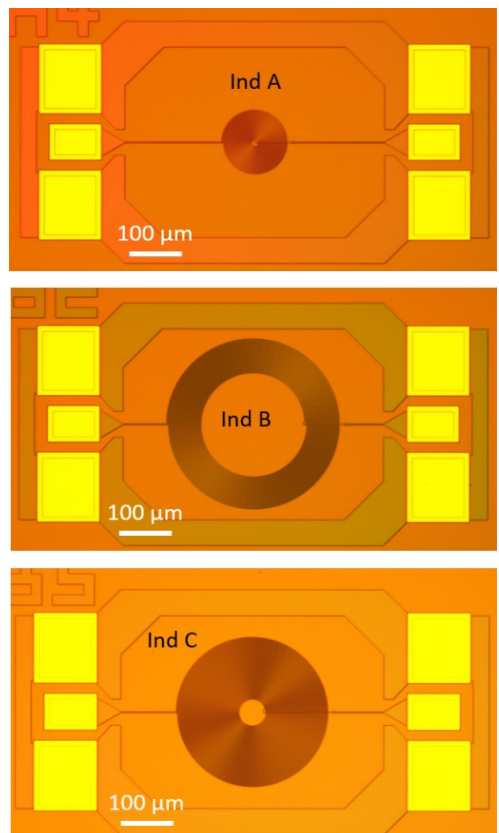

**Figure 23 - Optical micrographs of Ind A, Ind B ad Ind C.**